\documentclass[conference]{IEEEtran}

\IEEEoverridecommandlockouts
\usepackage{cite}
\usepackage{amsmath,amssymb,amsfonts}
\usepackage{graphicx}
\usepackage{textcomp}
\usepackage{xcolor}
\usepackage{booktabs}
\usepackage{array}
\usepackage{url}
\usepackage{hyperref}
\usepackage{microtype}

\hypersetup{
    colorlinks=true,
    linkcolor=black,
    citecolor=black,
    urlcolor=blue
}

\def\BibTeX{{\rm B\kern-.05em{\sc i\kern-.025em b}\kern-.08em
    T\kern-.1667em\lower.7ex\hbox{E}\kern-.125emX}}

\begin{document}

\title{Cybercrime Victimization Among Young Adult Males Aged 18--20:\\
A Post-Pandemic Analysis of Converging Risk Factors}

\author{
\IEEEauthorblockN{Anveeksh Mahesh Rao}
\IEEEauthorblockA{
\textit{Khoury College of Computer Sciences}\\
\textit{Northeastern University}\\
Boston, MA, USA\\
rao.anv@northeastern.edu\\
ORCID: 0000-0005-4791-1789
}
}

\maketitle

\begin{abstract}
Cybercrime victimization among young adult males aged 18--20 has become an
increasingly urgent public safety concern in the post-pandemic digital
environment. From 2022 to 2024, individuals aged 20--29 submitted 191,787
complaints to the FBI Internet Crime Complaint Center (IC3), reporting
combined losses of more than \$1.28 billion. Although this population
represents a substantial share of cybercrime victims, the 18--20 male
sub-cohort remains insufficiently examined as a distinct demographic group
within cybercrime victimization research. This study presents an original
risk factor analysis and theoretical synthesis, representing the first
integration of criminological, neurological, and behavioral evidence for this
specific demographic sub-cohort. Drawing on FBI IC3 and FTC Consumer Sentinel
Network data from 2022--2024 alongside European cybersecurity threat
intelligence from ENISA, the study develops a unified risk profile centered
on three intersecting vulnerability factors: a guardianship gap created by the
transition into unsupervised digital independence, heightened behavioral
exposure to online risk, and reduced impulse regulation associated with
ongoing prefrontal cortex development. The findings show a 49.7\% increase in
reported losses among the 20--29 age group between 2023 and 2024 and identify
financial sextortion, phishing, task scams, in-game currency fraud, and dark
web grooming as major attack typologies exploiting this risk convergence. The
study offers implications for targeted cybersecurity awareness campaigns,
digital literacy education, and policy interventions designed for this
high-risk demographic.
\end{abstract}


\section{Introduction}

Cybercrime has become one of the most widespread and consequential threats
affecting young adults in the post-pandemic digital era. Between 2022 and
2024, individuals aged 20--29 filed 191,787 complaints with the FBI Internet
Crime Complaint Center (IC3), reporting cumulative financial losses exceeding
\$1.28 billion \cite{ic3_2022,ic3_2023,ic3_2024}. Although IC3 reporting
groups victims into a broader 20--29 category, existing scholarship suggests
that the 18--20 sub-cohort displays especially elevated behavioral risk
patterns and vulnerability indicators \cite{nasi2015,griffith2023}. Federal
Trade Commission (FTC) data further demonstrates that young adults aged 20--29
reported losing money in 44\% of fraud cases, nearly twice the rate observed
among older populations, highlighting the financial susceptibility of this
group \cite{ftc2024}.

Gender is also a significant factor in cybercrime victimization among young
adults. N\"{a}si et al. \cite{nasi2015} found that males were 1.70 times more
likely than females to experience cybercrime victimization
($\text{OR}=1.70,\ p=0.000$), with particularly large gender gaps observed
in Germany (8.9\% male vs.\ 3.1\% female) and the United Kingdom (8.8\% male
vs.\ 5.9\% female). Griffith et al. \cite{griffith2023} further show that
online gaming, dating application use, visits to sexually explicit websites,
and public sharing of personal information increase exposure to motivated
offenders. These behaviors align closely with attack types that
disproportionately affect young men, including financial sextortion, phishing,
task scams, in-game currency fraud, and online grooming \cite{ray2025}.

The 18--20 age range represents a particularly acute period of vulnerability
for three reasons. First, it marks the transition from supervised adolescence
to independent digital life, producing a sudden reduction in external
guardianship \cite{griffith2023}. Second, several high-risk online behaviors
peak during this period, while cybersecurity awareness and protective habits
remain underdeveloped. Third, longitudinal fMRI evidence indicates that
ventrolateral prefrontal cortex (VLPFC) activation, which is associated with
impulse control and risk evaluation, continues to stabilize into the early
twenties \cite{qu2015}. As a result, young adult males in this age range face
a distinctive convergence of social, behavioral, and neurological risk factors.

Despite extensive research on cybercrime, adolescent risk-taking, and online
victimization, no prior study has fully integrated criminological, behavioral,
and neurological evidence into a unified risk profile focused specifically on
males aged 18--20. This study addresses that gap by examining how lifestyle
exposure, reduced guardianship, and developmental decision-making
vulnerabilities interact to shape cybercrime victimization risk in the
post-pandemic period.

The remainder of this paper is organized as follows.
Section~\ref{sec:lit} reviews relevant literature.
Section~\ref{sec:theory} presents the theoretical framework.
Section~\ref{sec:method} describes the methodology.
Section~\ref{sec:results} presents the findings.
Section~\ref{sec:discussion} discusses implications.
Section~\ref{sec:prevention} outlines prevention and policy recommendations.
Section~\ref{sec:conclusion} concludes the paper.

\section{Literature Review}
\label{sec:lit}

\subsection{Demographic Predictors of Cybercrime Victimization}

N\"{a}si et al. \cite{nasi2015} provide one of the most comprehensive
demographic analyses of cybercrime victimization among young people, using a
four-country sample from Finland, the United States, Germany, and the United
Kingdom ($n = 3{,}506$, aged 15--30). Their logistic regression analysis
found that male gender significantly predicted cybercrime victimization
($\text{OR}=1.70,\ p=0.000$), while younger age was also associated with
increased risk. Additional predictors included urban residence, unemployment,
and not living with parents, all of which suggest reduced social or
supervisory guardianship. The most commonly reported cybercrime types included
slander and threats of violence, with males disproportionately affected by
defamation and threat-based victimization. These patterns are consistent with
more recent forms of financially motivated sextortion and intimidation-based
online abuse documented in subsequent literature.

A key limitation of N\"{a}si et al. \cite{nasi2015} relevant to this study is
that the sample spans ages 15--30 without further disaggregation, preventing
direct statistical isolation of the 18--20 sub-cohort. This study addresses
that limitation by using the N\"{a}si et al. findings as a demographic anchor
while applying developmental and behavioral evidence to narrow the
vulnerability argument to the 18--20 window specifically.

\subsection{Routine Activity Theory and Behavioral Risk Factors}

Griffith et al. \cite{griffith2023} apply Lifestyle Routine Activity Theory
(LRAT) to cyber-victimization among young adults aged 18--25. Their findings
support the relevance of LRAT's three core elements in online contexts:
motivated offenders, suitable targets, and absent capable guardians. Online
dating emerged as a highly significant exposure variable across multiple
cybercrime types. Public disclosure of personal information and visits to
sexually explicit websites were associated with increased victimization risk
for obscenity, stalking, and bullying. Conversely, protective behaviors such
as using privacy settings and changing passwords reduced victimization
likelihood. These findings suggest that young adult males aged 18--20 may be
especially vulnerable when high online exposure coincides with weak protective
behaviors at precisely the moment parental oversight is removed.

\subsection{Financial Sextortion and Gender-Specific Targeting}

Ray and Henry \cite{ray2025} document the rapid global rise of financial
sextortion, identifying young males aged 18--24 as a frequently targeted
group by transnational organized crime networks. Offenders typically adopt
false female identities on social media platforms including Instagram and
Snapchat to establish trust, encourage sexual image exchange, and then deploy
threats of exposure to demand payment. These attacks can unfold rapidly,
sometimes within two weeks of initial contact. Because victims often
experience shame, embarrassment, and fear of social consequences, financial
sextortion is heavily underreported. As a result, official IC3 and FTC figures
likely underestimate the true scope of victimization within this demographic.
Ray and Henry \cite{ray2025} explicitly identify a gap in gender-differentiated
sextortion research, calling for studies examining how gender interacts with
age and developmental stage to shape victimization vulnerability --- a call
this study addresses directly.

\subsection{Neurological Foundations of Adolescent Risk-Taking}

Qu et al. \cite{qu2015} provide important developmental neuroscience evidence
linking prefrontal cortex development to adolescent risk-taking in the first
longitudinal fMRI study of this kind. Their study found that ventrolateral
prefrontal cortex (VLPFC) activation declines measurably during adolescence,
with greater VLPFC decline directly associated with greater real-world
risk-taking behavior. Because the VLPFC governs impulse control and rational
decision-making, continued neurological development into the early twenties
helps explain why individuals aged 18--20 remain vulnerable to emotionally
charged and reward-based social engineering tactics. This evidence supports
the view that cybercrime prevention strategies for young adults must account
not only for knowledge deficits but also for developmental limitations in
risk evaluation that information alone cannot overcome.

\subsection{Synthesis}

The four literature streams converge on a coherent but previously
unarticulated explanation for cybercrime vulnerability among young adult
males. N\"{a}si et al. \cite{nasi2015} establish the demographic pattern,
Griffith et al. \cite{griffith2023} identify the behavioral mechanisms, Ray
and Henry \cite{ray2025} describe the attack landscape, and Qu et al.
\cite{qu2015} explain the neurological basis for heightened risk-taking.
However, these findings have not previously been integrated into a single
unified risk model focused specifically on males aged 18--20. This study
synthesizes these perspectives to explain why this group occupies a
distinctive and underexamined position within cybercrime victimization
research.

\section{Theoretical Framework}
\label{sec:theory}

\subsection{Routine Activity Theory}

Routine Activity Theory (RAT), developed by Cohen and Felson \cite{cohen1979},
argues that crime occurs when three conditions converge: a motivated offender,
a suitable target, and the absence of a capable guardian. Choi \cite{choi2008}
extended this framework to cybercrime, demonstrating that online behaviors
function as digital equivalents of physical routine activities that bring
potential victims into proximity with motivated offenders.

\subsection{LRAT Applied to the 18--20 Male Sub-Cohort}

In this study, LRAT is applied to the 18--20 male sub-cohort across three
dimensions. \textit{Motivated offenders} are represented by organized criminal
networks and individual perpetrators who deliberately target young men through
social media, gaming platforms, dating applications, and encrypted
communication spaces \cite{ray2025}. \textit{Target suitability} is reflected
in behaviors such as dating app use, sexually explicit website visits, public
personal information sharing, and engagement in online gaming communities
\cite{griffith2023}. \textit{Absent guardianship} encompasses the decline of
parental oversight after age 18, weak cybersecurity habits, limited use of
privacy settings, and underdeveloped internal self-regulation.

\subsection{Internal Neurological Guardianship}

This study extends Routine Activity Theory by introducing the concept of
\textit{internal neurological guardianship} --- a dimension not previously
theorized within the RAT framework. In online social engineering contexts, the
ability to pause, evaluate risk, resist emotional manipulation, and avoid
impulsive action functions as a fundamental form of internal protection. The
prefrontal cortex serves as the neurological substrate of this internal
guardian. Because the VLPFC systems involved in impulse control and risk
assessment continue developing into the early twenties \cite{qu2015}, males
aged 18--20 experience a temporary but measurable reduction in this internal
guardianship capacity. This neurological vulnerability amplifies the effects
of external guardianship gaps and risky online routines, creating a
three-dimensional guardianship failure unique to this developmental window.

\section{Methodology}
\label{sec:method}

\subsection{Research Design}

This study uses a mixed secondary data analysis and systematic literature
review design. No primary data were collected; therefore, Institutional Review
Board (IRB) approval was not required. The analysis combines quantitative
cybercrime reporting data with qualitative synthesis of peer-reviewed research
to develop an integrated risk profile for males aged 18--20 during the
post-pandemic period from 2022 to 2024.

\subsection{Dataset Sources}

Three publicly available government datasets were analyzed.

\textbf{FBI IC3 Annual Reports (2022--2024):} Age-stratified complaint counts
and financial losses for the 20--29 age bracket were extracted from Appendix B
of each annual report \cite{ic3_2022,ic3_2023,ic3_2024}. As IC3 does not
disaggregate below the 20--29 bracket, literature-derived evidence is used
to support the 18--20 sub-cohort argument throughout the analysis.

\textbf{FTC Consumer Sentinel Network Data Book (2024):} Fraud report counts,
total losses, median losses, and fraud loss rates for individuals aged 20--29
were extracted to cross-validate FBI IC3 findings using an independent federal
data source \cite{ftc2024}.

\textbf{ENISA Threat Landscape Report (2024):} Used to provide international
context regarding phishing, social engineering, and emerging cyber threat
trends relevant to the target demographic \cite{enisa2024}.

\subsection{Systematic Literature Review Protocol}

A systematic literature review was conducted following PRISMA principles
using IEEE Xplore, ACM Digital Library, PubMed, Google Scholar, and SSRN.

\textit{Search terms} included: ``cybercrime victimization'' AND ``young
adults''; ``sextortion'' AND ``male'' AND ``18--25''; ``routine activity
theory'' AND ``cyber''; ``prefrontal cortex'' AND ``risk behavior'' AND
``adolescent''; ``online fraud'' AND ``gender'' AND ``youth.''

\textit{Inclusion criteria:} Peer-reviewed empirical articles or scoping
reviews published between 2008 and 2025; English language; focused on
cybercrime victimization, behavioral risk factors, or developmental
neuroscience relevant to young adults.

\textit{Exclusion criteria:} Studies focused exclusively on minors under 16;
non-cyber offenses; purely qualitative accounts without measurable findings.

Four primary sources were selected: N\"{a}si et al. \cite{nasi2015}, Griffith
et al. \cite{griffith2023}, Ray and Henry \cite{ray2025}, and Qu et al.
\cite{qu2015}.

\subsection{Limitations}

Four primary limitations are acknowledged. First, IC3 and FTC data cannot be
disaggregated below the 20--29 age bracket, preventing direct statistical
isolation of the 18--20 sub-cohort; the demographic argument therefore relies
on convergent literature evidence. Second, sextortion is substantially
underreported, meaning official figures represent conservative lower-bound
estimates \cite{ray2025}. Third, neurological evidence is applied
inferentially to online victimization contexts \cite{qu2015}. Fourth, the
use of English-language and US/EU-centered datasets limits global
generalizability.

\section{Results}
\label{sec:results}

\subsection{Three-Year IC3 Victimization Trend}

Table~\ref{tab:ic3} presents complaint counts and reported losses for the
20--29 age group extracted from FBI IC3 annual reports for 2022 through 2024.

\begin{table}[t]
\caption{FBI IC3 Complaint Data --- Age Group 20--29 (2022--2024)}
\label{tab:ic3}
\centering
\renewcommand{\arraystretch}{1.2}
\begin{tabular}{lrrr}
\hline
\textbf{Year} & \textbf{Complaints} & \textbf{Losses} &
\textbf{Change} \\
\hline
2022 & 57,978  & \$383.1M & ---       \\
2023 & 62,410  & \$360.7M & $-$5.8\%  \\
2024 & 71,399  & \$540.1M & +49.7\%   \\
\hline
\textbf{Total} & \textbf{191,787} & \textbf{\$1.28B} & +41.0\% \\
\hline
\multicolumn{4}{l}{\footnotesize Source: FBI IC3
\cite{ic3_2022,ic3_2023,ic3_2024}}
\end{tabular}
\end{table}

The data show a consistent increase in complaint volume year-on-year
(+7.6\% from 2022 to 2023; +14.4\% from 2023 to 2024). Reported losses
declined marginally in 2023 before increasing sharply by 49.7\% in 2024,
producing a cumulative three-year total of \$1.28 billion. This inflection
point is consistent with broader evidence of increasingly organized,
scalable, and psychologically sophisticated fraud operations, including
AI-assisted sextortion and social engineering campaigns targeting young
adult males \cite{ray2025}. The non-linear loss pattern suggests a
qualitative change in attacker capability between 2023 and 2024, rather
than a simple volume increase.

\subsection{FTC Consumer Sentinel Findings}

FTC Consumer Sentinel data independently reinforce the financial
vulnerability of young adults. In 2024, individuals aged 20--29 filed
155,346 fraud reports and reported \$430 million in total losses, with a
median individual loss of \$417 \cite{ftc2024}. Most notably, this group
reported losing money in 44\% of fraud cases --- nearly double the rate of
those aged 70--79 (24\%) and those aged 80 and above (21\%). This finding
directly challenges the common assumption that older adults are the most
fraud-susceptible demographic. Young adults are substantially more likely to
be successfully defrauded on any given contact, a pattern consistent with the
behavioral and neurological risk profile identified in this study.

\subsection{Synthesized Risk Factor Profile}

Table~\ref{tab:risk} presents the risk factor profile for males aged 18--20
organized by LRAT element, synthesized across the four primary literature
sources.

\begin{table}[t]
\caption{Risk Factor Profile --- Males Aged 18--20 (LRAT Framework)}
\label{tab:risk}
\centering
\renewcommand{\arraystretch}{1.2}
\begin{tabular}{p{1.4cm}p{4.5cm}p{1.2cm}}
\hline
\textbf{LRAT} & \textbf{Risk Factor} & \textbf{Ref.} \\
\hline
Motivated Offender &
Organized networks targeting young males via social media catfishing &
\cite{ray2025} \\
Motivated Offender &
High offender reach through social media, dating apps, and youth platforms &
\cite{ray2025} \\
Target Suitability &
Dating application use during emerging adulthood &
\cite{griffith2023} \\
Target Suitability &
Visits to sexually explicit websites &
\cite{griffith2023} \\
Target Suitability &
Public sharing of personal information &
\cite{griffith2023} \\
Target Suitability &
Male gender as significant victimization predictor
($\text{OR}=1.70,\ p=0.000$) &
\cite{nasi2015} \\
Absent Guardian &
Transition from supervised adolescence to independent digital life &
\cite{griffith2023} \\
Absent Guardian &
Limited privacy settings use and weak password practices &
\cite{griffith2023} \\
Absent Guardian &
Developing prefrontal impulse regulation (VLPFC immaturity) &
\cite{qu2015} \\
Absent Guardian &
Heightened reward sensitivity and risk-taking susceptibility &
\cite{qu2015} \\
\hline
\end{tabular}
\end{table}

The evidence supports a convergent risk profile for males aged 18--20. This
group faces motivated offenders who deliberately exploit youth-oriented
platforms, displays elevated target suitability due to peak-risk online
behaviors, and experiences multiple forms of absent guardianship including
reduced parental supervision, weak cybersecurity practices, and incomplete
neurological development. Together these factors create a distinctive
vulnerability structure in which young adult males are frequently exposed to
offenders, are behaviorally attractive targets, and often lack both external
and internal protective mechanisms simultaneously.

\subsection{Attack Typologies}

Six attack typologies appear especially relevant to males aged 18--20 based
on the synthesis of IC3 data and literature findings:

\begin{enumerate}
\item \textbf{Financial Sextortion:} The most gender-specific and emotionally
coercive attack type, exploiting sexual curiosity, dating behavior, and fear
of public humiliation. Blackmail threats can emerge within two weeks of first
contact \cite{ray2025}.

\item \textbf{Phishing and Spoofing:} High-volume attacks exploiting weak
password hygiene, limited cybersecurity awareness, and absent protective
habits \cite{ic3_2022,ic3_2023,ic3_2024}.

\item \textbf{Task Scams:} Fraud schemes targeting young adults entering
financial independence and seeking online income opportunities.

\item \textbf{Fake Investment Schemes:} Reward-driven scams exploiting
financial aspiration, limited investment experience, and heightened
neurological reward sensitivity \cite{qu2015}.

\item \textbf{In-Game Currency Fraud:} Gaming-related fraud exploiting
trust, competition, and reduced critical evaluation within immersive digital
environments \cite{griffith2023}.

\item \textbf{Online Grooming and Dark Web Recruitment:} Severe manipulation
exploiting social isolation and vulnerability during the transition into
adulthood.
\end{enumerate}

\section{Discussion}
\label{sec:discussion}

\subsection{The Convergent Vulnerability Profile}

The findings indicate that cybercrime victimization among young adult males
is not simply the result of poor awareness or careless behavior. It reflects
a convergence of developmental, behavioral, and environmental vulnerabilities.
The 49.7\% increase in IC3 losses from 2023 to 2024, combined with the FTC
finding that young adults lost money in 44\% of fraud reports, indicates that
traditional awareness campaigns are insufficient for this group. Neurological
evidence \cite{qu2015} suggests that risk information is less effective at
producing protective behavioral change in 18--20 year old males than in older
adults, because compromised VLPFC processing means impulsive responses to
social engineering stimuli are harder to override through deliberation alone.

\subsection{Behavioral-Neurological Amplification Dynamic}

A central contribution of this study is the identification of a
\textit{behavioral-neurological amplification dynamic}. The behavioral risks
identified by Griffith et al. \cite{griffith2023} do not operate
independently from the neurological patterns described by Qu et al.
\cite{qu2015}. Ongoing development in impulse control and reward evaluation
may help explain why some young adults engage in the very online behaviors
that increase their target suitability. Neurological vulnerability intensifies
behavioral exposure, while behavioral exposure creates more opportunities for
social engineering. This feedback loop is not captured by any single-discipline
approach and represents the primary theoretical contribution of this study.

\subsection{Post-Pandemic Context}

The post-pandemic period further amplifies these risks. Young adults
experienced increased screen time, expanded social media dependence, greater
online gaming engagement, and more frequent online income-seeking behavior
following COVID-19. These shifts placed young men in more frequent contact
with the precise digital environments where cybercriminals operate. At the
same time, the growth of AI-assisted fraud tools and organized transnational
sextortion networks has lowered the cost and increased the scale of attacks
targeting this demographic, consistent with the 49.7\% loss spike observed
in 2024.

\section{Prevention and Policy Implications}
\label{sec:prevention}

\subsection{Developmentally Calibrated Intervention Design}

Effective prevention must be developmentally calibrated rather than purely
informational. Three design principles follow from this study's findings.

First, platforms should adopt \textit{friction-based safety design}, including
delay prompts before sharing intimate images, stronger default privacy
settings, real-time impersonation warnings, and identity-verification alerts
in high-risk interactions. These mechanisms reduce reliance on spontaneous
user judgment during emotionally charged moments where VLPFC regulation is
least effective.

Second, interventions should be \textit{developmentally timed}, occurring at
ages 17--18 within secondary and post-secondary digital literacy curricula,
at the transition point immediately preceding the guardianship vacuum. Content
should address financial sextortion, phishing, gaming-related fraud, task
scams, and investment fraud using realistic scenarios tailored to young adult
behavior.

Third, prevention strategies must account for \textit{gender-specific barriers
to reporting}. Young men targeted by sextortion frequently avoid seeking help
due to shame and fear of social consequences \cite{ray2025}. Reporting systems
should be confidential, nonjudgmental, and designed to reduce stigma.

\subsection{Policy Recommendations}

Three policy recommendations follow from these findings. First, federal
cybercrime datasets including IC3 should adopt a dedicated 18--20 reporting
bracket to allow more precise demographic analysis and evidence-based resource
allocation. Second, social media and dating platforms should implement
design-level safety features targeting sextortion and organized social
engineering, including image-sharing friction mechanisms and organized crime
detection algorithms. Third, national digital literacy curricula should
incorporate cybercrime prevention content calibrated to the specific attack
typologies and behavioral risk patterns of the 18--20 male demographic
identified in this study.

\section{Conclusion}
\label{sec:conclusion}

This study develops the first unified risk profile for cybercrime
victimization among males aged 18--20 in the post-pandemic digital landscape.
By synthesizing FBI IC3 data, FTC fraud reports, ENISA threat intelligence,
and peer-reviewed criminological, behavioral, and neuroscientific literature,
the study shows that this demographic faces a unique convergence of
vulnerabilities: reduced guardianship during the transition to adulthood,
heightened exposure to risky online environments, and ongoing neurological
development affecting impulse control and risk evaluation.

The proposed behavioral-neurological amplification dynamic explains why
awareness-based prevention alone is insufficient. Young adult males are not
merely uninformed users; they are developmentally situated in a period where
reward sensitivity, social pressure, online exposure, and reduced oversight
combine to increase cybercrime susceptibility in ways that information alone
cannot address. More effective interventions must therefore be
platform-integrated, developmentally timed, and sensitive to the gendered
stigma that prevents reporting.

Future research should pursue finer-grained federal cybercrime reporting
brackets, longitudinal studies linking online victimization to neurological
development measures, and empirical evaluations of platform-based safety
interventions against the attack typologies identified in this study.

\section*{Author Statement}

This paper presents original research conceived, designed, and independently
conducted by the author. The research framework, analytical approach,
theoretical contributions, and conclusions represent the original intellectual
work of the author. Secondary data analysis was performed independently using
publicly available government datasets from the FBI Internet Crime Complaint
Center, the Federal Trade Commission Consumer Sentinel Network, and the
European Union Agency for Cybersecurity. All cited references were
independently verified by the author against their original published sources
prior to submission. AI writing tools were used for drafting and editing
assistance during manuscript preparation. No AI-generated references,
fabricated citations, or hallucinated content are present in this work. The
author accepts full responsibility for the accuracy, integrity, and
originality of all content submitted.


\end{document}